\ifdefined\luatexversion
  \RequirePackage{luatex85}
\fi

\documentclass[preprintsapa,article,accept,moreauthors]{Definitions/mdpi}

\usepackage{amsmath,amssymb,amsfonts}
\usepackage{booktabs}
\usepackage{multirow}
\usepackage{array}
\usepackage{tabularx}
\newcolumntype{L}[1]{>{\raggedright\arraybackslash}p{#1}}
\newcolumntype{P}[1]{>{\centering\arraybackslash}p{#1}}

\firstpage{1}
\pubvolume{1}
\issuenum{1}
\articlenumber{0}
\pubyear{2026}
\copyrightyear{2026}
\datereceived{ }
\daterevised{ }
\dateaccepted{ }
\datepublished{ }

\Title{Interaction Effects Between Learner Characteristics and Dialogue Format in LLM- and TTS-Generated Dialogue-Based Lessons}

\Author{Fumie Watanabe $^{1,}$*, Gendo Kumoi $^{2}$, Tota Suko $^{3}$, Takashi Ishida $^{4}$, Yuko Kuma $^{5}$, Manabu Kobayashi $^{1}$ and Shigeichi Hirasawa $^{1}$}

\AuthorNames{Fumie Watanabe, Gendo Kumoi, Tota Suko, Takashi Ishida, Yuko Kuma, Manabu Kobayashi and Shigeichi Hirasawa}

\address{%
$^{1}$ \quad Center for Data Science, Waseda University, Tokyo 169-8050, Japan\\
$^{2}$ \quad School of Engineering, Nagaoka University of Technology, Niigata 940-2188, Japan\\
$^{3}$ \quad Faculty of Social Sciences, Waseda University, Tokyo 169-8050, Japan\\
$^{4}$ \quad Faculty of Economics, Takasaki City University of Economics, Gunma 370-0801, Japan\\
$^{5}$ \quad Faculty of Architecture and Civil Engineering, Kyushu Sangyo University, Fukuoka 813-8503, Japan}

\corres{Correspondence: watanabe\_fumie@aoni.waseda.jp}

\abstract{Large language models and text-to-speech technology enable low-cost production of dialogue-based lessons: videos in which two speakers converse. The speakers' roles can vary, but little is known about whether all learners respond alike to such differences. In a repeated-measures design, 222 first-year high school students viewed lessons in three formats: a teacher and a student (TS), two students (SS), and two teachers (TT). Within the aptitude--treatment interaction framework, linear mixed-effects models tested whether associations between learner characteristics and outcomes differed by format. The learner characteristics were experiential learning style---seeking challenging, new experiences (CE factor) and learning by reflecting on experience (RCE factor)---and critical thinking disposition; the outcomes were motivation based on the ARCS model, self-reported learning outcomes, and overall evaluation. For motivation, both factors interacted with format: the CE factor's association differed between TS and TT, and the RCE factor's slope was significantly positive in TS and SS and significantly larger in TS than TT. For learning outcomes, no interaction was detected, but critical thinking disposition was positively associated with them. For overall evaluation, again no interaction was detected, but TT was rated lowest. Given small-to-medium interaction effects, these findings are preliminary evidence to inform personalized learning design.}


\begin{document}

\section{Introduction}

\subsection{Background}

Generative artificial intelligence (AI), in particular large language models (LLMs) combined with text-to-speech (TTS) technology, has made it possible to produce lesson videos at low cost. One such form, and the focus of this study, is the dialogue-based lesson: a video in which two speakers converse about the lesson content. Because the speakers' roles can be set when the script is generated~\citep{do2025paige,kumoi2026dialoguetts}, such lessons can take the form of, for example, a dialogue between a teacher and a student (TS), between two students (SS), or between two teachers (TT). Generative AI thus makes it practical to produce lessons in multiple dialogue formats.

These formats differ from one another in who the speakers are and how they interact: for example, whether one of the speakers takes the role of a learner, and whether one speaker explains while the other asks questions or both explain to each other as peers. Research on learning from observed dialogue suggests that such differences matter: students learned more from tutorial dialogues than from monologues, with the tutee's contributions eliciting constructive engagement from observers~\citep{chi2017dialogue}. Moreover, higher-performing students were reported to learn more from dialogue videos in which the tutee made errors than from the corresponding monologue videos, whereas no such difference was found among lower-performing students~\citep{ding2021dialogue}. The same dialogue format may therefore work differently for different learners, which raises a design question for dialogue-based lessons: do all learners respond to these formats in the same way?

This question bears on personalization, a prominent expectation in research on AI in education (AIEd). What remains unclear is which aspects of learners should guide how instruction is tailored, and on what theoretical grounds~\citep{bernacki2021pl,holmes2022aied}. Aptitude--treatment interaction (ATI) offers one theoretical framework for addressing this~\citep{cronbach1977ati}. ATI holds that the outcomes of instruction are jointly determined by learner characteristics (aptitudes) and instructional methods (treatments), so that the effectiveness of a given method can differ across learners. Treating dialogue format as the treatment, the question becomes whether the associations between learner characteristics and the outcomes of a lesson differ across dialogue formats.

This study focuses on two learner characteristics. The first is experiential learning style, grounded in D.~A.~Kolb's~\citeyearpar{kolb1984} experiential learning theory, which captures individual differences in how learners engage with and process experience---for example, the extent to which they seek new experiences or learn by reflecting on them~\citep{ikejiri2021}. Because a dialogue-based lesson is itself an experience for the learner, these differences may shape how learners respond to it. The second is critical thinking disposition, a tendency to logically evaluate and judge presented information~\citep{kusumi2025cti}, which may account for differences in how learners process the content of a discussion between speakers.

Learners' responses to a lesson have more than one aspect, and these aspects do not necessarily align. In studies comparing AI-generated with conventionally produced instructional videos, the conventionally produced videos were rated higher in learning experience~\citep{netland2025aivideo} or social presence~\citep{xu2025aivideo}, but the videos rated higher did not also yield better learning: test performance did not differ significantly~\citep{netland2025aivideo}, and retention was higher for the AI-generated video~\citep{xu2025aivideo}. Netland et al.~\citeyearpar{netland2025aivideo} also found that learners' ratings were almost unrelated to their test performance. Judging lessons by learners' ratings alone may therefore lead to conclusions different from those based on their learning, so the two need to be examined separately. Accordingly, this study examines three outcome measures: motivation based on the ARCS model (\citealp{keller1987arcs}; hereafter, motivation), self-reported learning outcomes, and overall evaluation. Motivation captures learners' motivational response to a lesson in terms of attention, relevance, confidence, and satisfaction. The ARCS model is a method for improving the motivational appeal of instruction: it provides strategies for each of these four components and a systematic process for designing them into lessons~\citep{keller1987arcs}. This makes motivation a suitable outcome for a study examining the design of dialogue formats. Self-reported learning outcomes capture whether learners felt they understood the content. Overall evaluation captures how learners rate each lesson as a whole.

\subsection{Related Work}

\subsubsection{AI in Education and Personalization}\label{sec:aied-personalization}

Research on AIEd has grown across educational levels. A systematic review of AIEd in higher education identified applications in profiling and prediction, assessment and evaluation, adaptive systems and personalization, and intelligent tutoring systems~\citep{zawackirichter2019aied}, and a review of K-12 studies grouped affordances into pedagogy (e.g., personalized learning, gaming, and collaboration), administration, and subject content tools~\citep{crompton2024k12}. More broadly, Holmes and Tuomi~\citeyearpar{holmes2022aied} classified AIEd applications as student-, teacher-, and institution-focused, and distinguished the use of AI as a means of teaching and learning from teaching about AI itself (AI literacy). Since generative AI tools became widely available, studies of their use have accumulated across educational levels, from higher education~\citep{ogunleye2024genai} to K-12 education, including the secondary level. A systematic review of ChatGPT use in K-12 education found that the most frequently reported opportunities were personalized learning, differentiated instruction, and support for educators in generating assessments and learning tasks, and noted that ChatGPT could support educators in lesson planning and materials generation~\citep{zhang2024chatgptk12}. Similarly, a systematic review of generative AI in secondary education noted that it allows teachers to produce materials quickly and adapt them to learners~\citep{perezmontesdeoca2026genai}. The present study uses generative AI in this way, to produce lesson materials, and examines how learners with different characteristics respond to them. It thus concerns the use of AI as a means of teaching and learning rather than AI literacy.

Conceptual work has described how AIEd has changed. Ouyang and Jiao~\citeyearpar{ouyang2021paradigms} organized AIEd into three paradigms: AI-directed, in which learners follow system-defined learning paths; AI-supported, in which learners and systems interact; and AI-empowered, in which learners take agency in their own learning. In their position paper, they described the development trend of AIEd as moving toward learner agency and personalization.

Personalization is a prominent expectation in this literature, but its meaning and basis remain contested. Holmes and Tuomi~\citeyearpar{holmes2022aied} note that the meaning of personalized learning remains unclear, and regard the adaptive pathways through learning materials offered by much current AIEd as a limited interpretation of personalization. They also point out that the effects of learning interventions can depend on learners' prior knowledge. A systematic review of personalized learning found that designs frequently adapted to learners' prior knowledge, performance, and preferences or interests, and that this research tended to proceed without a priori theoretical conceptualization~\citep{bernacki2021pl}. A review of AIEd in higher education similarly reported a weak connection to theoretical pedagogical perspectives~\citep{zawackirichter2019aied}, and conceptual work has identified the same problem across systematic reviews of AIEd~\citep{ouyang2021paradigms}. Thus, even in the narrower, adaptive sense of personalization, what to adapt to, and on what theoretical grounds, remains an open design question.

One constraint on personalized and adaptive instruction has been the effort required to build its underlying models and materials. Holmes and Tuomi~\citeyearpar{holmes2022aied} observe that domain models for real-world settings are often difficult and expensive to define, and Crompton et al.~\citeyearpar{crompton2024k12} suggest that more complex AI systems could require more development time and expertise, and hence be more costly. Producing instructional videos has also been costly: presenting a human instructor entails higher financial and time costs, and reducing these costs has been a stated rationale for AI-generated instructors~\citep{pi2022voice}.

Generative AI may reduce part of this constraint. LLMs and TTS technology can generate lesson scripts and narrated videos at low production cost; D.~Zhang~\citeyearpar{zhang2025semiauto}, using a semi-automated system that combined Google Gemini-generated scripts with Amazon Polly TTS, estimated that AI-generated video could be produced faster than conventionally produced video, and in a two-course pilot study found no significant difference in quiz scores compared with a previous semester that used conventionally produced videos. Other studies comparing AI-generated with conventionally produced instructional videos have also examined learners' affective responses. In Xu et al.~\citeyearpar{xu2025aivideo}, total test scores and transfer did not differ significantly, whereas retention was significantly higher for the AI-generated video; motivation and satisfaction did not differ significantly, although the conventionally produced video was rated higher in social presence. Netland et al.~\citeyearpar{netland2025aivideo} likewise found no significant difference in test performance, but the conventionally produced videos were rated slightly higher in learning experience, and participants' ratings of their learning experience were almost unrelated to their test performance.

\subsubsection{Dialogue-Based Materials Using LLMs and TTS}\label{sec:dialogue-materials}

Research on the automated generation of dialogue-based educational content by combining LLMs and TTS technology has progressed rapidly in recent years. Kumoi et al.~\citeyearpar{kumoi2026dialoguetts} examined the educational potential of LLM- and TTS-generated dialogue-based lessons themselves. They proposed a system that semi-automatically generates such lessons as dialogues between a teacher (Expert) and a student (Novice), and they evaluated three formats---teacher video, single-speaker TTS, and dialogue TTS---with 245 first-year high school students. Of twelve pre-specified outcomes, nine did not differ significantly between dialogue TTS and single-speaker TTS; the dialogue TTS lesson was rated higher in learners' confidence in their own understanding but lower in audio quality and higher in extraneous cognitive load. The authors therefore describe the dialogue format as offering a trade-off rather than a general advantage. Nonetheless, 66.9\% of respondents chose the dialogue format as the ``most enjoyable'' way to learn, which the authors treat as a preference rather than evidence of effectiveness. Because the study used a fixed presentation order and different lesson content across formats, the authors position these results as observed patterns rather than causal effects of format.

In another study of AI-generated dialogue, Do et al.~\citeyearpar{do2025paige} developed PAIGE (Personalized AI-Generated Educational podcasts), which converts a textbook chapter into a dialogue script between a host and an expert via an LLM and delivers it as a podcast voiced by two TTS speakers. In a $3\times3$ experimental design with 180 university students, learners rated the podcasts as more enjoyable than textbook reading, and a version personalized to each learner's profile (e.g., major and interests) yielded higher learning outcomes in some subjects. This result suggests that dialogue-format AI audio content can be more enjoyable for learners than reading, and that the benefit of personalization for learning may depend on the subject.

Dialogue with an LLM has also been used for tutoring without TTS. Li et al.~\citeyearpar{li2024tutorly} built on cognitive apprenticeship theory to develop Tutorly, a system that turns programming videos into a one-on-one tutoring experience in which learners converse in text with an LLM-based mentor. In a within-subjects study, 16 participants completed both a Tutorly condition and a baseline condition in which they watched a video directly while using their own IDE and ChatGPT, and their test scores improved from a pre-test to a post-test administered after both conditions. Unlike the lessons reviewed above, in which learners listen to a generated dialogue, Tutorly has learners interact with the LLM directly.

\subsubsection{Learner Characteristics: Experiential Learning Style and Critical Thinking Disposition}\label{sec:learner-characteristics}

D.~A.~Kolb~\citeyearpar{kolb1984} proposed Experiential Learning Theory, which describes learning as a cyclical process across four modes: Concrete Experience, Reflective Observation, Abstract Conceptualization, and Active Experimentation. Individual learning style is characterized by the relative preference for each mode~\citep{kolbkolb2005}, which suggests that learners may respond differently to the same lesson format.

A.~Y.~Kolb and Kolb~\citeyearpar{kolbkolb2005} applied this theory to higher education and introduced the concept of ``learning space'' as a framework for understanding the interface between learners' styles and the educational environment. They describe learning style as a dynamic state, not a fixed trait, and as a position on continuous dimensions rather than a categorical type. They also note that a learner's position in the learning space reflects both individual disposition and characteristics of the learning environment. Experiential learning theory likewise holds that learning results from transactions between the person and the environment, which is compatible with the ATI view that learning outcomes depend on the combination of learner characteristics and instructional method~\citep{cronbach1977ati}. Within this view, lesson formats that differ in how they present information, such as stepwise teacher-led explanation and collaborative dialogue between peers, can be examined for how they relate to learners' ways of processing experience.

D.~A.~Kolb's four modes have also been measured with an experiential learning style scale developed in Japan~\citep{kimura2011}. In a survey of young workers using this scale, Ikejiri et al.~\citeyearpar{ikejiri2021} found that a two-factor structure was more appropriate than the four-factor structure, and Ikejiri et al.~\citeyearpar{ikejiri2022} named the two factors the concrete experience factor and the factor of active experimentation through reflective observation and abstract conceptualization. In this study, we refer to them as the CE (concrete experience) factor and the RCE (reflection, conceptualization, and experimentation) factor, respectively. The CE factor reflects a tendency to seek challenging, new experiences, and the RCE factor reflects a tendency to learn by reflecting on experience~\citep{ikejiri2021}.

Critical thinking disposition is another learner characteristic linked to learning engagement. Lv et al.~\citeyearpar{lv2022} conducted a longitudinal study of 654 students at Chinese higher vocational institutions and found a reciprocal positive relationship between critical thinking disposition and learning engagement. This finding suggests that learners with a strong critical thinking disposition are more likely to engage in active, reflective learning behavior.

Taken together, within the ATI framework, these two characteristics can be treated as aptitudes and dialogue format as the treatment, which makes it possible to examine whether the associations of each characteristic with the outcomes of a lesson differ across dialogue formats.

\subsection{Purpose of This Study}

The literature reviewed above leaves two gaps. First, comparisons of AI-generated videos have mainly contrasted them with conventionally produced videos, and studies of LLM- and TTS-generated dialogue have generally evaluated a single dialogue format or contrasted dialogue with monologue. Kumoi et al.~\citeyearpar{kumoi2026dialoguetts}, for example, compared a teacher--student dialogue with single-speaker TTS and found a trade-off rather than a general advantage. Comparisons of dialogue formats that differ in the speakers' roles, made within the same group of learners, therefore remain limited. Second, although the reviewed work suggests that experiential learning style and critical thinking disposition may shape how learners respond to a lesson, how these characteristics interact with dialogue format has not been examined.

This study therefore aims to examine, in an exploratory manner within the ATI framework, whether the associations between learner characteristics and the outcomes of LLM- and TTS-generated dialogue-based lessons differ across dialogue formats that differ in the speakers' roles. To this end, we set the following research question (RQ).

RQ: In LLM- and TTS-generated dialogue-based lessons, do the associations of learners' experiential learning style (CE factor, RCE factor) and critical thinking disposition with the outcomes of the lessons (motivation, learning outcomes, and overall evaluation) differ across dialogue formats (TS, SS, and TT)?

The main contributions of this study are as follows.

1) It provides empirical evidence on a research gap that has received limited attention to date: the interaction between learner characteristics and dialogue format in LLM- and TTS-generated dialogue-based lessons.

2) By applying the ATI framework, it offers a theoretical lens for examining how the relationship between learners' cognitive and affective characteristics and their responses differs across dialogue formats.

3) By examining how learner characteristics relate to three outcome measures---motivation, learning outcomes, and overall evaluation---across dialogue formats, it offers hypotheses to be tested in future research on the design of AI-generated instructional materials.

\begin{figure}[H]
\centering
\includegraphics[width=\textwidth]{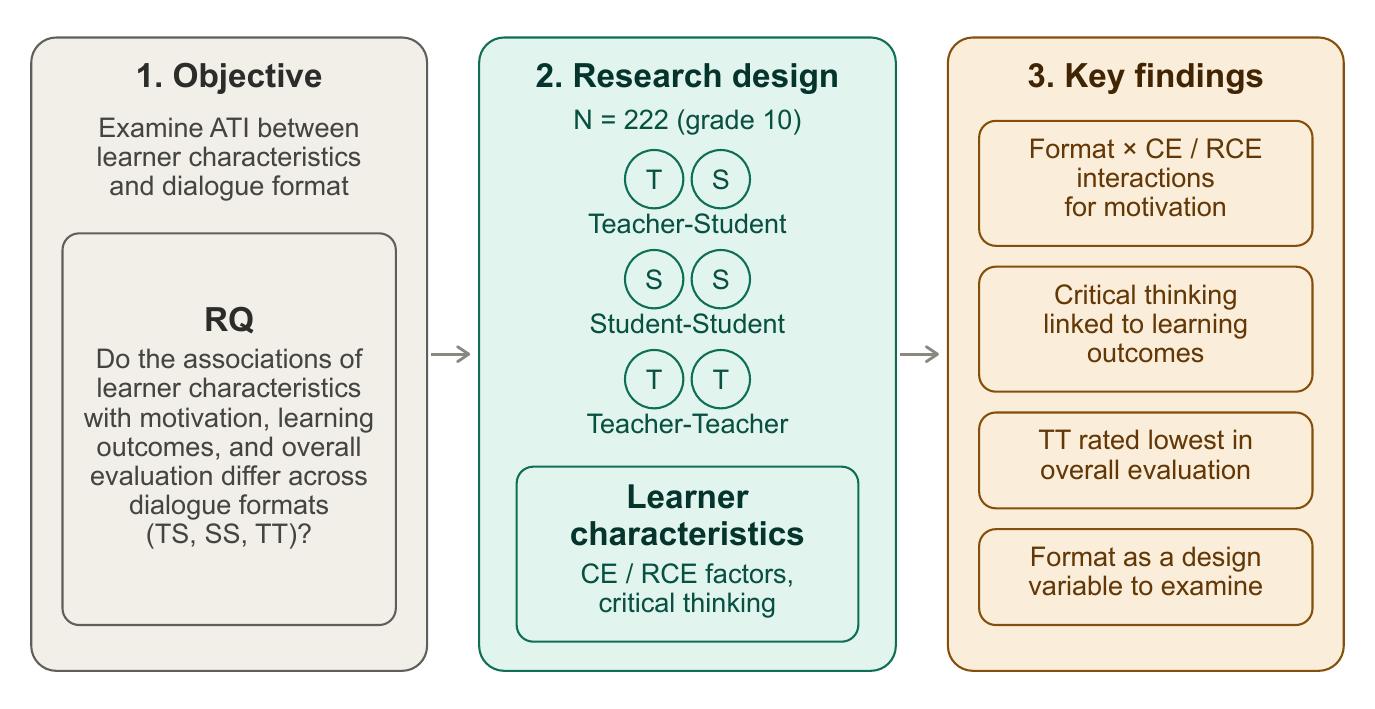}
\caption{Overview of this study (objective, research design, and key findings).}
\label{fig:overview}
\end{figure}

Figure~\ref{fig:overview} summarizes the objective, research design, and key findings of this study.

\section{Lessons}

\subsection{Overview of the Lessons}

The lessons in this study were conducted as part of ``General Inquiry Time'' at a public high school. They formed part of a data-driven career education program that used generative AI, and they covered content related to data science.

Part 1 (TS format, approximately 25 min) addressed ``what data science is and why it is attracting attention now.'' Part 2 (SS format, approximately 19 min) addressed ``what data science can do,'' using the analysis of convenience-store point-of-sale data, web recommendation systems, and sports statistics as examples. Part 3 (TT format, approximately 24 min) addressed ``data-driven decision making,'' covering the concept of evidence-based policy making, the relationship between descriptive analytics and prediction, and the principle that ``prediction is impossible without description,'' thereby emphasizing data literacy and basic summarization skills. Total viewing time was approximately 68 minutes.

\subsection{Structure of the Dialogue Formats}\label{sec:dialogue-formats}

The lessons consisted of the following three dialogue formats; the procedures for generating the scripts and synthesizing the speech are described below. In the TS format, a teacher role explained concepts while a student role, who was new to the topic, asked questions and restated the explanations in their own words, building understanding step by step. In the SS format, two student roles who were still learning the topic took turns as explainer and listener, expressing uncertainty and correcting each other's misunderstandings as they built understanding together as equal learning partners. In the TT format, an expert role explained specialized content in greater depth, while a facilitator role, who led the discussion as a knowledgeable host rather than as a learner, asked the expert role questions, periodically organized the points, and restated the expert role's explanations.

The number of dialogue turns was approximately 180 for the TS format, 190 for the SS format, and 75 for the TT format, with mean utterance lengths of approximately 40, 30, and 100 Japanese characters (excluding punctuation), respectively. That is, the TS and SS formats proceeded through frequent, short exchanges between the two speakers, whereas in the TT format the speakers took fewer turns, each consisting of a longer stretch of speech.

Dialogue scripts were generated using Claude 3.5 Sonnet (Anthropic). Prompts were designed separately for each format: an Expert-Novice prompt generated the Expert (teacher) and Novice (student) dialogue for the TS format, a Peer-Peer prompt generated the peer-to-peer dialogue for the SS format, and an Expert-Facilitator prompt generated the expert-facilitator dialogue for the TT format. Speech was synthesized using the Gemini TTS API (\texttt{gemini-2.5-pro-preview-tts}, google-genai SDK), with the same two prebuilt voices in all three formats. Enceladus was assigned to the speaker who mainly explained (the teacher role in the TS format and the expert role in the TT format) and Autonoe to the other speaker (the student role in the TS format and the facilitator role in the TT format); in the SS format, Enceladus and Autonoe were assigned to the two student roles, respectively. The three formats thus differed in the speakers' roles rather than in their voices. Only \texttt{response\_modalities=["AUDIO"]} was specified as a generation setting; sampling temperature and other parameters were left unset. Audio was output as 24~kHz, mono, 16-bit PCM WAV, and prosody was controlled with inline tags (e.g., \texttt{[short pause]}, \texttt{[medium pause]}, \texttt{[uhm]}). The generation scripts and prompt templates are publicly available in a GitHub repository (\url{https://github.com/moiku/gemini-tts-lecture-generation}), in which the TS, SS, and TT formats correspond to the Expert $\times$ Novice (EN), Peer $\times$ Peer (PP), and Expert $\times$ Facilitator (EF) patterns, respectively.

\subsection{Lesson Administration Procedure}\label{sec:lesson-procedure}

All students viewed the TS, SS, and TT formats sequentially within the same session (a within-subjects design). After viewing each format, students completed a survey about that format. After viewing and responding to all three formats, students rated each format in a comparative section, which provided the overall evaluation measure, and then wrote free-response comments. Because the lessons were administered simultaneously across all six classes during regular school hours, counterbalancing the presentation order was not feasible.

Lessons were viewed via projector in each classroom rather than in a single shared venue; all six participating classes (approximately 40 students each) viewed the lessons simultaneously in their own classrooms. No individual audio devices such as earphones were used, and students viewed the lessons under the classroom's ambient acoustic conditions. A single short break was provided during the viewing of the three formats.

\section{Materials and Methods}

\subsection{Participants, Timing, and Procedure}

An online questionnaire survey was administered to 222 first-year students at a public high school who took part in the lessons. Most participants had also taken part in the study by Kumoi et al.~\citeyearpar{kumoi2026dialoguetts}, conducted between May and July 2025. The two studies used different lessons and collected separate survey data, and no data from that study were used in the present analyses.

The survey was conducted on November 12, 2025, the day of the lessons, following the procedure described in Section~\ref{sec:lesson-procedure}. The questionnaire was distributed via Google Forms, and students responded using the iPad issued to each of them by the school.

\subsection{Survey Content}

\subsubsection{Survey of Learner Characteristics}\label{sec:experiential-learning-scale}

This study measured experiential learning style and critical thinking disposition as learner characteristics.

(1) Experiential learning style

Experiential learning style was measured using the 16-item experiential learning style scale developed by Kimura et al.~\citeyearpar{kimura2011}. Following Ikejiri et al.~\citeyearpar{ikejiri2021,ikejiri2022}, this study adopted the two-factor structure (CE factor: 4 items; RCE factor: 12 items). Responses were given on a 5-point scale (1: not at all, to 5: always).

(2) Critical thinking disposition

Critical thinking disposition was measured using the 10-item Critical Thinking attitude in Inquiry scale (CT-I) developed by Kusumi~\citeyearpar{kusumi2025cti}. This scale measures critical thinking disposition in high school students' inquiry-based learning and was judged to be well suited in content to the present sample. Responses were given on a 5-point scale (1: does not apply, to 5: applies).

\subsubsection{Survey of Responses to the Lessons}\label{sec:lesson-evaluation-survey}

For each dialogue-format lesson, the following were measured: three outcome measures, namely (1) motivation, (2) learning outcomes, and (3) overall evaluation, as well as (4) supplementary check items and (5) free-response items.

(1) Motivation

Motivation was measured with a 4-item scale based on Keller's~\citeyearpar{keller2010} ARCS model, with one item corresponding to each of Attention, Relevance, Confidence, and Satisfaction. Responses were given on a 5-point scale (1: strongly disagree, to 5: strongly agree).

(2) Learning outcomes

Learning outcomes were measured with a 2-item index: ``I understood the main points of the lesson well'' and ``I think I could explain the main content of the lesson to a friend.'' Responses were given on a 5-point scale (1: strongly disagree, to 5: strongly agree).

(3) Overall evaluation

Overall evaluation was designed to capture an impression of the dialogue format that was conceptually independent of motivation and learning outcomes. Under the instruction ``please rate each lesson on a 5-point scale,'' a single item (``overall rating'') asked for an overall evaluation of each dialogue format. Responses were given on a 5-point scale (1: very poor, to 5: very good).

(4) Supplementary check items

The dialogue formats differed in the number of turns and mean utterance length (Section~\ref{sec:dialogue-formats}). Because spoken information is transient, longer spoken segments can impose a heavier load on working memory~\citep{leahysweller2011}. To check whether learners' perceived difficulty differed across dialogue formats, two items were measured for each format: ``the content of the lesson itself was difficult'' and ``the way the lesson was conducted was hard to follow, making it difficult to concentrate on the content.'' Responses were given on a 5-point scale (1: strongly disagree, to 5: strongly agree).

(5) Free-response items

For each dialogue format, students were asked to provide free-response comments on areas for improvement.

\subsection{Analysis}

\subsubsection{Linear Mixed-Effects Models}

To examine the interaction effects of learner characteristics (CE factor, RCE factor, critical thinking disposition) and dialogue format (TS, SS, TT) on each outcome measure, we used linear mixed-effects models. Participant intercepts were entered as a random effect to account for individual differences, and dialogue format, learner characteristics, and their interaction terms were entered as fixed effects. Analyses were conducted using the lme4 and lmerTest packages in R. Learner-characteristic variables were grand-mean centered. The TS format served as the reference category for dialogue format.

Multicollinearity among learner-characteristic variables was assessed using the variance inflation factor (VIF). Differences across formats in the two supplementary check items were examined using the Friedman test ($\alpha = .05$).

Fixed effects were evaluated with Type III $F$ tests, with degrees of freedom approximated by the Satterthwaite method. For each outcome measure, this yielded seven tests: the main effect of dialogue format, the main effects of the three learner characteristics, and the three format $\times$ learner characteristic interactions. Because the analyses were exploratory, the seven tests in each model were treated as one family, following Cramer et al.~\citeyearpar{cramer2016multiplicity}, who noted that all effects in an exploratory multiway analysis of variance are tested jointly. Among the corrections they described, controlling the false discovery rate is suited to identifying candidate effects for later confirmatory testing, whereas controlling the familywise error rate guards more strictly against false positives at the cost of power. We therefore applied the Benjamini--Hochberg procedure~\citep{benjamini1995fdr} within each family to control the false discovery rate, and also report $p$ values adjusted with Holm's method, which controls the familywise error rate, to show which effects remained under this more conservative criterion. Adjusted $p$ values reported in the text are those from the Benjamini--Hochberg procedure unless Holm's method is specified. As an effect size, partial Cohen's $f$ for each effect was computed from the same analysis of variance with the \texttt{cohens\_f} function of the effectsize package.

For format $\times$ learner characteristic interactions that were significant after Benjamini--Hochberg correction, simple main effects (slopes) for each dialogue format were estimated from the full model that included all three characteristics, using the \texttt{emtrends} function of the emmeans package, and differences in slopes between dialogue formats were tested with Tukey-corrected pairwise comparisons. For outcome measures with a significant main effect of dialogue format and no significant format $\times$ learner characteristic interaction, differences between formats were tested with Tukey-corrected pairwise comparisons of the estimated marginal means obtained with the \texttt{emmeans} function. Degrees of freedom for the emmeans analyses were computed with the Kenward--Roger method. Because all learner-characteristic variables were measured on the same 5-point scale, they were not standardized, so that a one-point change in each variable directly indicates the expected difference in the outcome measures.

\subsubsection{Free-Response Analysis}

Free-response comments were analyzed through qualitative coding. First, for the TS-format responses, the lead coder assigned single-concept, noun-phrase labels based on the KJ method, an affinity-diagram approach to grouping qualitative data~\citep{sato2019qualitative}. These human-coded TS-format labels then served as few-shot examples for automatically labeling the SS- and TT-format responses with an LLM, and the lead coder reviewed and corrected all labels (human-in-the-loop).

To avoid the risk that labeling by a single coder (the lead coder) would be arbitrary, the TS-format comments were independently re-coded by the LLM in a separate session without reference to the original human-assigned labels (zero-shot re-classification), and agreement with the human-coded labels was assessed using Cohen's $\kappa$. Agreement was $\kappa = .952$ (near-perfect), supporting the reliability of the coding.

Within each format, labels that occurred only once were merged into an ``other'' category. Claude Sonnet 4.6 (Anthropic, as of June 2026) was used for both the few-shot and zero-shot classification.

\subsection{Ethical Considerations}

Data collection was conducted anonymously, without collecting any personally identifying information such as name, student ID, sex, or age. The study was conducted with the approval of the school principal, and the guardians had given comprehensive written consent to the school's provision of students' information, such as their interests and activity records, to third-party organizations approved by the principal. At the beginning of the questionnaire, students were informed that participation was voluntary, that they could withdraw at any time, that their responses would not affect their grades, that anonymized, statistically processed results might be published in academic outlets, and that the data would not be used for any purpose other than research; only students who agreed to these terms responded. Response data were recorded and analyzed in a form that precluded individual identification from the outset, and the survey was administered as part of ordinary educational activities.

\section{Results}

\subsection{Comparison of Perceived Difficulty Across Formats}

Because the number of turns and mean utterance length differed across dialogue formats, we first checked whether learners' perceived difficulty and clarity of the lesson differed across formats. A Friedman test comparing the three formats on the two supplementary check items showed no significant difference for either item (content difficulty: $\chi^2(2) = 1.62$, $p = .444$; clarity of delivery: $\chi^2(2) = 1.13$, $p = .568$; Kendall's $W < .01$ for both).

\subsection{Examination of the Measures}

\subsubsection{Reliability of the Experiential Learning Style Scale}

For the 16-item experiential learning style scale, the internal consistency of each factor in the two-factor structure adopted from Ikejiri et al.~\citeyearpar{ikejiri2021,ikejiri2022} was examined. The CE factor (4 items) yielded $\alpha = .750$, and the RCE factor (12 items) yielded $\alpha = .840$, both exceeding the acceptable threshold ($\alpha \geq .70$).

\subsubsection{Factor Structure and Reliability of the Critical Thinking Disposition Scale}

For the 10-item critical thinking disposition scale, we conducted exploratory factor analysis using maximum likelihood estimation and adopted a one-factor solution, because the first eigenvalue (3.44) was more than three times the second (1.08) and the one-factor solution was the most interpretable. A confirmatory factor analysis of this one-factor model (robust maximum likelihood estimation) showed acceptable fit (robust CFI $= .967$, robust TLI $= .957$, robust RMSEA $= .040$, SRMR $= .045$), and internal consistency was $\alpha = .782$. In this study, critical thinking disposition thus functioned as a single, integrated dimension rather than as differentiated subfactors.

\subsubsection{Reliability of the Outcome Measures}

Internal consistency of the motivation, learning outcomes, and overall evaluation measures was examined separately for each dialogue format (Table~\ref{tab:alpha}).

\begin{table}[H]
\caption{Cronbach's $\alpha$ coefficients for the outcome measures}
\label{tab:alpha}
\centering
\begin{tabular}{lccc}
\toprule
Scale (items) & TS & SS & TT \\
\midrule
Motivation (4 items) & .820 & .837 & .894 \\
Learning outcomes (2 items)   & .718 & .676 & .734 \\
Overall evaluation (1 item)   & ---   & ---   & ---   \\
\bottomrule
\end{tabular}
\end{table}

Motivation showed Cronbach's $\alpha = .82$--$.89$ across all formats, indicating sufficient internal consistency. Learning outcomes showed $\alpha = .68$--$.73$, which was judged acceptable for a 2-item scale. Overall evaluation, being a single item, was not subject to a reliability coefficient.

\subsubsection{Multicollinearity Check}

VIF values for the learner-characteristic variables (CE factor, RCE factor, critical thinking disposition) were all below 5 (maximum VIF $= 2.16$), indicating no problematic multicollinearity. Correlations among the variables were $r = .58$ between the CE and RCE factors, $r = .43$ between the CE factor and critical thinking disposition, and $r = .65$ between the RCE factor and critical thinking disposition.

\subsection{Descriptive Statistics and Correlation Analysis}

Descriptive statistics for all variables are shown in Table~\ref{tab:desc}.

\begin{table}[H]
\caption{Descriptive statistics for all variables ($N = 222$)}
\label{tab:desc}
\centering
\begin{tabular}{lcccc}
\toprule
Variable & Mean & SD & Min & Max \\
\midrule
\multicolumn{5}{l}{Learner characteristics} \\
\quad CE factor           & 3.49 & 0.73 & 1.75 & 5.00 \\
\quad RCE factor   & 3.72 & 0.53 & 2.17 & 5.00 \\
\quad Critical thinking disposition       & 3.59 & 0.56 & 1.80 & 5.00 \\
\midrule
\multicolumn{5}{l}{Motivation} \\
\quad TS & 3.42 & 0.70 & 1.00 & 5.00 \\
\quad SS & 3.49 & 0.71 & 1.00 & 5.00 \\
\quad TT & 3.52 & 0.75 & 1.00 & 5.00 \\
\midrule
\multicolumn{5}{l}{Learning outcomes} \\
\quad TS & 3.39 & 0.77 & 1.00 & 5.00 \\
\quad SS & 3.46 & 0.73 & 1.00 & 5.00 \\
\quad TT & 3.45 & 0.80 & 1.00 & 5.00 \\
\midrule
\multicolumn{5}{l}{Overall evaluation} \\
\quad TS & 3.54 & 0.85 & 1.00 & 5.00 \\
\quad SS & 3.54 & 0.96 & 1.00 & 5.00 \\
\quad TT & 3.41 & 0.82 & 1.00 & 5.00 \\
\bottomrule
\end{tabular}
\end{table}

Correlation analysis between learner characteristics and outcome measures showed that the RCE factor was significantly and positively correlated with motivation across all three formats ($r = .23$--$.28$, $p < .001$) and with learning outcomes across all three formats ($r = .20$--$.31$, all $p < .01$). Critical thinking disposition was likewise significantly and positively correlated with motivation ($r = .26$--$.29$, $p < .001$) and learning outcomes ($r = .31$--$.37$, $p < .001$). The CE factor was significantly and positively correlated only with motivation in the TT format ($r = .15$, $p = .022$) and with learning outcomes in the SS format ($r = .16$, $p = .020$). For overall evaluation, the only significant correlations were those with critical thinking disposition ($r = .24$, $p < .001$) and the RCE factor ($r = .15$, $p = .027$) in the TT format.

\subsection{Examination of Interactions Using Linear Mixed-Effects Models}

Table~\ref{tab:f-tests} summarizes the Type III $F$ tests of the fixed effects for the three outcome measures.

\begin{table}[H]
\caption{Type III $F$ tests of the fixed effects in the linear mixed-effects models}
\label{tab:f-tests}
\centering
\small
\begin{tabularx}{\textwidth}{l>{\raggedright\arraybackslash}Xrrrrr}
\toprule
 & & & & \multicolumn{2}{c}{Adjusted $p$} & \\
\cmidrule(lr){5-6}
Outcome measure & Effect & $F$ & $p$ & BH & Holm & Partial $f$ \\
\midrule
Motivation & Dialogue format & 7.13 & $<.001$ & \textbf{.006} & \textbf{.006} & .18 \\
 & CE factor & 0.96 & .327 & .327 & .367 & .07 \\
 & RCE factor & 3.64 & .058 & .081 & .173 & .13 \\
 & Critical thinking disposition & 5.94 & .016 & \textbf{.027} & .062 & .17 \\
 & Format $\times$ CE factor & 6.29 & .002 & \textbf{.007} & \textbf{.012} & .17 \\
 & Format $\times$ RCE factor & 5.10 & .006 & \textbf{.015} & \textbf{.032} & .15 \\
 & Format $\times$ Critical thinking disposition & 1.70 & .183 & .214 & .367 & .09 \\
\midrule
Learning outcomes & Dialogue format & 2.24 & .108 & .237 & .604 & .10 \\
 & CE factor & 1.46 & .228 & .311 & .683 & .08 \\
 & RCE factor & 1.24 & .267 & .311 & .683 & .08 \\
 & Critical thinking disposition & 18.71 & $<.001$ & $\mathbf{<.001}$ & $\mathbf{<.001}$ & .29 \\
 & Format $\times$ CE factor & 2.01 & .135 & .237 & .604 & .10 \\
 & Format $\times$ RCE factor & 2.31 & .101 & .237 & .604 & .10 \\
 & Format $\times$ Critical thinking disposition & 0.08 & .925 & .925 & .925 & .02 \\
\midrule
Overall evaluation & Dialogue format & 5.21 & .006 & \textbf{.041} & \textbf{.041} & .15 \\
 & CE factor & 0.01 & .915 & .970 & 1.000 & .01 \\
 & RCE factor & 0.00 & .970 & .970 & 1.000 & .00 \\
 & Critical thinking disposition & 3.31 & .070 & .164 & .351 & .12 \\
 & Format $\times$ CE factor & 1.13 & .324 & .567 & 1.000 & .07 \\
 & Format $\times$ RCE factor & 0.38 & .683 & .957 & 1.000 & .04 \\
 & Format $\times$ Critical thinking disposition & 3.05 & .049 & .164 & .291 & .12 \\
\bottomrule
\multicolumn{7}{p{\dimexpr\textwidth-2\tabcolsep\relax}}{\footnotesize Note. Type III $F$ tests from the full linear mixed-effects models (Tables~\ref{tab:lmm-arcs}, \ref{tab:lmm-outcome}, and \ref{tab:lmm-eval}), with degrees of freedom approximated by the Satterthwaite method: $F(2, 436)$ for dialogue format and the interactions, and $F(1, 218)$ for the learner characteristics. Adjusted $p$ values were computed within each model across its seven tests with the Benjamini--Hochberg (BH) procedure and Holm's method. Adjusted $p$ values below .05 are shown in bold. Partial $f$ = partial Cohen's $f$.} \\
\end{tabularx}
\end{table}

\subsubsection{Effects on Motivation}

Table~\ref{tab:lmm-arcs} shows the fixed-effect estimates of the full model for motivation.

\begin{table}[H]
\caption{Fixed effects of the linear mixed-effects model for motivation}
\label{tab:lmm-arcs}
\centering
\small
\begin{tabularx}{\textwidth}{Xrrrrl}
\toprule
Term & Coefficient ($b$) & SE & $t$ & $p$ & \\
\midrule
Intercept & 3.423 & 0.046 & 73.75 & $<.001$ & *** \\
\multicolumn{6}{l}{Dialogue format (reference: TS)} \\
\quad SS & 0.068 & 0.027 & 2.50 & $.013$ & * \\
\quad TT & 0.100 & 0.027 & 3.70 & $<.001$ & *** \\
\multicolumn{6}{l}{Learner characteristics (slopes in the TS format)} \\
\quad CE factor             & $-0.151$ & 0.078 & $-1.93$ & $.055$ & \\
\quad RCE factor     &  0.341 & 0.130 &  2.63 & $.009$ & ** \\
\quad Critical thinking disposition         &  0.222 & 0.110 &  2.02 & $.044$ & * \\
\multicolumn{6}{l}{Interaction terms} \\
\quad SS $\times$ CE factor         &  0.074 & 0.046 &  1.61 & $.109$ & \\
\quad TT $\times$ CE factor         &  0.162 & 0.046 &  3.54 & $<.001$ & *** \\
\quad SS $\times$ RCE factor & $-0.083$ & 0.076 & $-1.10$ & $.273$ & \\
\quad TT $\times$ RCE factor & $-0.238$ & 0.076 & $-3.15$ & $.002$ & ** \\
\quad SS $\times$ Critical thinking disposition     & $-0.008$ & 0.064 & $-0.12$ & $.907$ & \\
\quad TT $\times$ Critical thinking disposition     &  0.099 & 0.064 &  1.54 & $.125$ & \\
\bottomrule
\multicolumn{6}{p{\dimexpr\textwidth-2\tabcolsep\relax}}{\footnotesize Marginal $R^2$ (fixed effects) $= .095$; Conditional $R^2$ (fixed and random effects) $= .846$} \\
\multicolumn{6}{p{\dimexpr\textwidth-2\tabcolsep\relax}}{\footnotesize Note. * $p < .05$, ** $p < .01$, *** $p < .001$} \\
\multicolumn{6}{p{\dimexpr\textwidth-2\tabcolsep\relax}}{\footnotesize Random effects: participant intercept variance $= 0.397$, residual variance $= 0.081$, $N = 666$ (222 participants $\times$ 3 formats)} \\
\end{tabularx}
\end{table}

The format $\times$ CE and format $\times$ RCE interactions were significant after Benjamini--Hochberg correction (adjusted $p = .007$ and $.015$, respectively) and remained significant with Holm's method (adjusted $p = .012$ and $.032$; Table~\ref{tab:f-tests}). The format $\times$ critical thinking disposition interaction was not significant (adjusted $p = .214$). The effect sizes of the two significant interactions were small to medium (partial $f = .17$ and $.15$, respectively; \citealp{Cohen1988}). The fixed effects accounted for Marginal $R^2 = .095$ of the variance. The main effect of dialogue format was also significant ($F(2, 436) = 7.13$, adjusted $p = .006$ with both correction methods; Table~\ref{tab:f-tests}). Because dialogue format interacted with the CE and RCE factors, this main effect was not followed by pairwise comparisons; instead, the interactions were examined through the simple main effects (slopes) reported below.

\begin{table}[H]
\caption{Simple main effects (slopes and 95\% CIs) of the CE factor on motivation}
\label{tab:simple-effects-ce}
\centering
\begin{tabular}{L{1.8cm}P{1.6cm}P{3.4cm}}
\toprule
Dialogue format & Slope & 95\% CI \\
\midrule
TS & $-$0.151 & [$-$0.306, 0.003] \\
SS & $-$0.078 & [$-$0.232, 0.077] \\
TT & 0.011 & [$-$0.144, 0.166] \\
\bottomrule
\multicolumn{3}{p{6.8cm}}{\footnotesize Note. None of the slopes were significant (all 95\% CIs included zero).} \\
\end{tabular}
\end{table}

\begin{table}[H]
\caption{Pairwise differences in slopes of the CE factor across dialogue formats (Tukey-corrected)}
\label{tab:simple-effects-ce-diff}
\centering
\begin{tabular}{L{2.8cm}P{1.4cm}P{1.2cm}P{1.4cm}}
\toprule
Comparison & Slope diff. & $t$ & $p$ \\
\midrule
TS $-$ SS & $-0.074$ & $-1.61$ & $.243$ \\
TS $-$ TT & $-0.162$ & $-3.54$ & $.001$\textsuperscript{**} \\
SS $-$ TT & $-0.089$ & $-1.93$ & $.131$ \\
\bottomrule
\multicolumn{4}{p{6.8cm}}{\footnotesize Note. \textsuperscript{**}$p < .01$ (Tukey-adjusted). The slope difference is the slope of the format on the left of the ``Comparison'' column minus that on the right; the sign indicates the direction of the difference.} \\
\end{tabular}
\end{table}

\begin{table}[H]
\caption{Simple main effects (slopes and 95\% CIs) of the RCE factor on motivation}
\label{tab:simple-effects-rce}
\centering
\begin{tabular}{L{1.8cm}P{1.6cm}P{3.4cm}}
\toprule
Dialogue format & Slope & 95\% CI \\
\midrule
TS & 0.341 & [0.085, 0.596] \\
SS & 0.257 & [0.002, 0.513] \\
TT & 0.102 & [$-$0.153, 0.358] \\
\bottomrule
\multicolumn{3}{p{6.8cm}}{\footnotesize Note. The slopes were significant for the TS and SS formats ($p < .05$) but not for the TT format (95\% CI included zero).} \\
\end{tabular}
\end{table}

\begin{table}[H]
\caption{Pairwise differences in slopes of the RCE factor across dialogue formats (Tukey-corrected)}
\label{tab:simple-effects-rce-diff}
\centering
\begin{tabular}{L{2.8cm}P{1.4cm}P{1.2cm}P{1.4cm}}
\toprule
Comparison & Slope diff. & $t$ & $p$ \\
\midrule
TS $-$ SS & 0.083 & $1.10$ & $.516$ \\
TS $-$ TT & 0.239 & $3.15$ & $.005$\textsuperscript{**} \\
SS $-$ TT & 0.155 & $2.05$ & $.102$ \\
\bottomrule
\multicolumn{4}{p{6.8cm}}{\footnotesize Note. \textsuperscript{**}$p < .01$ (Tukey-adjusted). The slope difference is the slope of the format on the left of the ``Comparison'' column minus that on the right; the sign indicates the direction of the difference.} \\
\end{tabular}
\end{table}

For the two learner characteristics showing a significant interaction, we examined the simple main effects (slopes) for each dialogue format. For the CE factor (Tables~\ref{tab:simple-effects-ce} and \ref{tab:simple-effects-ce-diff}), none of the slopes were significant individually, but the slopes differed significantly between the TS and TT formats ($t(436) = -3.54$, $p = .001$).

For the RCE factor (Tables~\ref{tab:simple-effects-rce} and \ref{tab:simple-effects-rce-diff}), the slope was significantly positive in the TS format (0.341) and the SS format (0.257), but not significant in the TT format (0.102). The slope in the TS format was significantly larger than that in the TT format ($t(436) = 3.15$, $p = .005$).

For critical thinking disposition, whose interaction with dialogue format was not significant, the main effect was significant after Benjamini--Hochberg correction ($F(1, 218) = 5.94$, adjusted $p = .027$) and showed a trend toward significance with Holm's method (adjusted $p = .062$); learners with higher critical thinking disposition reported higher motivation (Table~\ref{tab:f-tests}).

\subsubsection{Effects on Learning Outcomes}

Table~\ref{tab:lmm-outcome} shows the fixed-effect estimates of the full model including all variables.

\begin{table}[H]
\caption{Fixed effects of the linear mixed-effects model for learning outcomes}
\label{tab:lmm-outcome}
\centering
\small
\begin{tabularx}{\textwidth}{Xrrrrl}
\toprule
Term & Coefficient ($b$) & SE & $t$ & $p$ & \\
\midrule
Intercept & 3.387 & 0.048 & 70.31 & $<.001$ & *** \\
\multicolumn{6}{l}{Dialogue format (reference: TS)} \\
\quad SS &  0.074 & 0.037 &  1.99 & $.048$ & * \\
\quad TT &  0.061 & 0.037 &  1.62 & $.105$ & \\
\multicolumn{6}{l}{Learner characteristics (slopes in the TS format)} \\
\quad CE factor             & $-0.137$ & 0.082 & $-1.68$ & $.093$ & \\
\quad RCE factor     &  0.264 & 0.135 &  1.96 & $.051$ & \\
\quad Critical thinking disposition         &  0.425 & 0.114 &  3.72 & $<.001$ & *** \\
\multicolumn{6}{l}{Interaction terms} \\
\quad SS $\times$ CE factor         &  0.121 & 0.063 &  1.91 & $.057$ & \\
\quad TT $\times$ CE factor         &  0.027 & 0.063 &  0.42 & $.673$ & \\
\quad SS $\times$ RCE factor & $-0.195$ & 0.105 & $-1.86$ & $.063$ & \\
\quad TT $\times$ RCE factor & $-0.195$ & 0.105 & $-1.86$ & $.064$ & \\
\quad SS $\times$ Critical thinking disposition     &  0.015 & 0.089 &  0.17 & $.868$ & \\
\quad TT $\times$ Critical thinking disposition     &  0.035 & 0.089 &  0.39 & $.694$ & \\
\bottomrule
\multicolumn{6}{p{\dimexpr\textwidth-2\tabcolsep\relax}}{\footnotesize Marginal $R^2$ (fixed effects) $= .129$; Conditional $R^2$ (fixed and random effects) $= .737$} \\
\multicolumn{6}{p{\dimexpr\textwidth-2\tabcolsep\relax}}{\footnotesize Note. * $p < .05$, ** $p < .01$, *** $p < .001$} \\
\multicolumn{6}{p{\dimexpr\textwidth-2\tabcolsep\relax}}{\footnotesize Random effects: participant intercept variance $= 0.360$, residual variance $= 0.156$, $N = 666$ (222 participants $\times$ 3 formats)} \\
\end{tabularx}
\end{table}

For learning outcomes, none of the format $\times$ learner characteristic interactions was significant after correction (minimum adjusted $p = .237$; Table~\ref{tab:f-tests}). Among the main effects of learner characteristics, critical thinking disposition was significant with both correction methods ($F(1, 218) = 18.71$, adjusted $p < .001$), with higher disposition associated with higher learning outcomes, whereas the CE and RCE factors were not (adjusted $p = .311$). The main effect of dialogue format was not significant (adjusted $p = .237$). The fixed effects accounted for Marginal $R^2 = .129$ of the variance.

\subsubsection{Effects on Overall Evaluation}

Table~\ref{tab:lmm-eval} shows the fixed-effect estimates of the full model including all variables.

\begin{table}[H]
\caption{Fixed effects of the linear mixed-effects model for overall evaluation}
\label{tab:lmm-eval}
\centering
\small
\begin{tabularx}{\textwidth}{Xrrrrl}
\toprule
Term & Coefficient ($b$) & SE & $t$ & $p$ & \\
\midrule
Intercept & 3.541 & 0.059 & 60.28 & $<.001$ & *** \\
\multicolumn{6}{l}{Dialogue format (reference: TS)} \\
\quad SS &  0.000 & 0.045 &  0.00 & $1.000$ & \\
\quad TT & $-0.126$ & 0.045 & $-2.79$ & $.005$ & ** \\
\multicolumn{6}{l}{Learner characteristics (slopes in the TS format)} \\
\quad CE factor             &  0.022 & 0.099 &  0.22 & $.829$ & \\
\quad RCE factor     & $-0.069$ & 0.164 & $-0.42$ & $.675$ & \\
\quad Critical thinking disposition         &  0.205 & 0.139 &  1.47 & $.142$ & \\
\multicolumn{6}{l}{Interaction terms} \\
\quad SS $\times$ CE factor         &  0.039 & 0.076 &  0.50 & $.614$ & \\
\quad TT $\times$ CE factor         & $-0.074$ & 0.076 & $-0.97$ & $.330$ & \\
\quad SS $\times$ RCE factor &  0.100 & 0.126 &  0.79 & $.429$ & \\
\quad TT $\times$ RCE factor &  0.090 & 0.126 &  0.71 & $.475$ & \\
\quad SS $\times$ Critical thinking disposition     & $-0.097$ & 0.107 & $-0.91$ & $.364$ & \\
\quad TT $\times$ Critical thinking disposition     &  0.164 & 0.107 &  1.53 & $.126$ & \\
\bottomrule
\multicolumn{6}{p{\dimexpr\textwidth-2\tabcolsep\relax}}{\footnotesize Marginal $R^2$ (fixed effects) $= .030$; Conditional $R^2$ (fixed and random effects) $= .714$} \\
\multicolumn{6}{p{\dimexpr\textwidth-2\tabcolsep\relax}}{\footnotesize Note. * $p < .05$, ** $p < .01$, *** $p < .001$} \\
\multicolumn{6}{p{\dimexpr\textwidth-2\tabcolsep\relax}}{\footnotesize Random effects: participant intercept variance $= 0.540$, residual variance $= 0.226$, $N = 666$ (222 participants $\times$ 3 formats)} \\
\end{tabularx}
\end{table}

For overall evaluation, none of the format $\times$ learner characteristic interactions was significant after correction (minimum adjusted $p = .164$; Table~\ref{tab:f-tests}). None of the main effects of learner characteristics was significant (minimum adjusted $p = .164$). The main effect of dialogue format was significant with both correction methods ($F(2, 436) = 5.21$, adjusted $p = .041$). Tukey-corrected pairwise comparisons showed that the TT format received a lower rating than both the TS and SS formats (difference $= 0.126$, $t(436) = 2.79$, $p = .015$ for both comparisons), whereas the TS and SS formats did not differ significantly ($p = 1.000$). The fixed effects accounted for Marginal $R^2 = .030$ of the variance.

\subsection{Results of the Free-Response Analysis}

Free-response comments on areas for improvement, which were optional, were provided by 38 students (17.1\%) for the TS format, 44 students (19.8\%) for the SS format, and 46 students (20.7\%) for the TT format. Coding these responses yielded 19 distinct labels. Within each format, labels that occurred only once were merged into an ``other'' category; Table~\ref{tab:qual-def} shows the definitions of the 14 labels that remained in at least one format.

\begin{table}[H]
\caption{Definitions of the improvement-point labels}
\label{tab:qual-def}
\centering
\small
\begin{tabularx}{\textwidth}{l>{\raggedright\arraybackslash}X}
\toprule
Label & Definition \\
\midrule
Unnatural conversation content & Comments about a sense of unnaturalness in the flow, structure, setting, or content of the conversation \\
Unnatural manner of speaking & Comments that the tone, manner, or reactions in the speech felt odd or unnatural \\
Stiffness of speech & Comments that the speech was stiff and explanatory, making it feel unapproachable \\
Unnatural intonation & Comments that the intonation or pronunciation accent felt odd \\
Unnatural voice quality & Comments about unnaturalness or inconsistency in voice color, voice quality, or perceived age \\
Unnatural backchanneling & Comments that the backchannel responses were formulaic or repetitive and sounded unnatural \\
Monotonous conversation & Comments that the conversation lacked variation or enjoyment and felt tedious or monotonous \\
Unclear speaker roles & Comments that it was hard to tell the position or role of each speaker \\
Dissatisfaction with the student role's questions & Comments that the content or depth of the student role's questions was insufficient or inappropriate \\
Dissatisfaction with speech rate & Comments that the speech was too fast to follow \\
Redundant explanation & Comments that explanations were too long or roundabout and lacked conciseness \\
Difficulty of content & Comments that the lesson content was difficult to understand \\
Abundance of difficult vocabulary & Comments that the words and expressions used were difficult to understand \\
Unapproachability of the format & Comments that the dialogue format felt unfamiliar, making the content hard to take in \\
\bottomrule
\end{tabularx}
\end{table}

Tables~\ref{tab:qual-ts}, \ref{tab:qual-ss}, and \ref{tab:qual-tt} show the frequency of the improvement-point labels for each dialogue format.

\begin{table}[H]
\caption{Frequency of improvement-point labels for the TS format}
\label{tab:qual-ts}
\centering
\begin{tabular}{lr}
\toprule
Label & Frequency \\
\midrule
Unnatural conversation content & 6 \\
Unnatural manner of speaking & 5 \\
Dissatisfaction with the student role's questions & 5 \\
Dissatisfaction with speech rate & 4 \\
Redundant explanation & 4 \\
Unnatural backchanneling & 4 \\
Unnatural intonation & 2 \\
Unclear speaker roles & 2 \\
Unnatural voice quality & 2 \\
Monotonous conversation & 2 \\
Other & 4 \\
\bottomrule
\end{tabular}
\end{table}

\begin{table}[H]
\caption{Frequency of improvement-point labels for the SS format}
\label{tab:qual-ss}
\centering
\begin{tabular}{lr}
\toprule
Label & Frequency \\
\midrule
Unnatural manner of speaking & 14 \\
Unclear speaker roles & 5 \\
Unnatural conversation content & 4 \\
Unnatural intonation & 4 \\
Unnatural voice quality & 4 \\
Redundant explanation & 4 \\
Dissatisfaction with speech rate & 4 \\
Unapproachability of the format & 2 \\
Other & 5 \\
\bottomrule
\end{tabular}
\end{table}

\begin{table}[H]
\caption{Frequency of improvement-point labels for the TT format}
\label{tab:qual-tt}
\centering
\begin{tabular}{lr}
\toprule
Label & Frequency \\
\midrule
Stiffness of speech & 8 \\
Unclear speaker roles & 6 \\
Dissatisfaction with speech rate & 5 \\
Unnatural intonation & 4 \\
Difficulty of content & 3 \\
Redundant explanation & 3 \\
Abundance of difficult vocabulary & 3 \\
Unnatural voice quality & 2 \\
Monotonous conversation & 2 \\
Unapproachability of the format & 2 \\
Other & 4 \\
\bottomrule
\end{tabular}
\end{table}

In the TT format, ``stiffness of speech'' (frequency 8) was the most common label, and ``difficulty of content'' and ``abundance of difficult vocabulary'' also emerged as distinct labels. In the SS format, ``unnatural manner of speaking'' (frequency 14) was the most common label. In the TS format, ``unnatural conversation content'' (frequency 6) was the most common label.

\section{Discussion}

Within the ATI framework, the RQ of this study was whether the associations of experiential learning style (CE and RCE factors) and critical thinking disposition with the outcomes of LLM- and TTS-generated dialogue-based lessons differ across dialogue formats. Below, we discuss the results for each outcome measure in turn, first considering the interactions, which address the RQ, and then the main effects, which are not part of the RQ but help to interpret how learners responded to the lessons.

\subsection{Motivation}

For motivation, the format $\times$ CE and format $\times$ RCE interactions were significant. Examination of the simple main effects showed that, for the CE factor, none of the slopes was significant in any format, although the slopes differed between the TS and TT formats; that is, how the CE factor was associated with motivation differed between these formats. Because no individual slope was significant, however, the present data do not indicate whether the CE factor was positively or negatively associated with motivation in either format, or in which format learners high in the CE factor were more motivated. We therefore regard this interaction as a finding to be examined further in a later confirmatory study.

For the RCE factor, by contrast, learners higher in this factor reported higher motivation in the TS and SS formats, and the slope in the TS format was significantly larger than that in the TT format. This pattern can be interpreted through the ATI framework. Ikejiri et al.~\citeyearpar{ikejiri2021} interpreted the RCE factor as reflection-based actions taken after experiences. In the TS format, the student role, who was new to the topic, restated the teacher role's explanations in their own words, and understanding was built step by step through these exchanges. This may have allowed learners high in the RCE factor to reflect on and conceptualize the content alongside the student role. In a study in which pairs of university students watched instructional videos together, pairs who watched a tutorial dialogue between a tutor and a tutee engaged more constructively than pairs who watched a lecture-style monologue, and the tutee's contributions were suggested to elicit this engagement~\citep{chi2017dialogue}. In the TT format, the facilitator role also restated the explanations, but as a knowledgeable host organizing the points of the discussion. Such restatements present the content in an already organized form, which may have left comparatively little room for learners' own reflection. This may explain why the slope of the RCE factor was larger in the TS format than in the TT format.

Critical thinking disposition, whose interaction with format was not significant, was positively associated with motivation, although this main effect was significant only under the Benjamini--Hochberg criterion and showed a trend under Holm's method. The CT-I scale used in this study covers an emphasis on logic and evidence, inquisitiveness toward multiple perspectives, and a tendency to relate what is learned to other subjects and everyday life~\citep{kusumi2025cti}. In dialogue-based lessons, in which ideas are developed through an exchange between speakers, such a disposition may help learners follow the reasoning, consider the perspective offered by each speaker, and connect the content to what they already know, which may in turn have made the lessons more engaging for them.

\subsection{Learning Outcomes}

For learning outcomes, no significant interaction between learner characteristics and dialogue format was detected, and the CE and RCE factors showed no significant main effects. Critical thinking disposition, by contrast, showed a positive main effect that was significant under both correction methods. As with motivation, this disposition may help learners follow the reasoning of the discussion and connect it to what they already know, which may in turn have given them a sense of having understood the lessons.

\subsection{Overall Evaluation}

For overall evaluation, no interaction with learner characteristics was detected, and only the main effect of format was significant: the TT format was rated lower than both the TS and SS formats. One possible explanation for the lower evaluation of the TT format is that a lesson consisting of an exchange between two adult speakers---an expert and a facilitator---with no student role differs from first-year high school students' everyday learning experience, and this unfamiliarity may have lowered the overall impression of the lesson. The free-response comments are consistent with this interpretation: for the TT format, stiffness of speech was the most frequently noted issue, and difficulty of content and abundance of difficult vocabulary also emerged as distinct labels.

\subsection{Implications for Educational Practice}

These results suggest that the way learner characteristics relate to motivation can differ across dialogue formats. Dialogue format may therefore be worth considering as a design variable, together with learner characteristics, in LLM- and TTS-generated dialogue-based lessons. This bears on the call for theoretically grounded personalization~\citep{bernacki2021pl}. Rather than presupposing that formats should be matched to types of learners, this study examined whether such interactions exist in the first place, a step that may need to precede attempts at adaptation. However, the interactions were small to medium in size. In addition, they were detected for motivation but not for learning outcomes or overall evaluation. The implications below should therefore be read as hypotheses for future testing, not as established guidance.

In the present data, the associations of the CE and RCE factors with motivation differed between the TS and TT formats. These data do not show whether adapting the format to such patterns would improve motivation or learning. Testing this directly is a next step. For example, multiple dialogue formats could be provided in a learning management system, and four conditions compared: a fixed-format condition; an assigned condition, in which the system selects a format based on a brief pre-assessment of learner characteristics; a recommended condition, in which the system suggests a format based on the same pre-assessment and learners make the final decision; and a self-selected condition, in which learners choose a format themselves. The assigned and recommended conditions thus differ only in who decides. The three conditions other than the fixed-format one differ in how much agency they leave to learners. Ouyang and Jiao~\citeyearpar{ouyang2021paradigms} used this dimension to distinguish three paradigms of AIEd: AI-directed, AI-supported, and AI-empowered. Comparing them with the fixed-format condition would show whether adapting the format to learner characteristics improves motivation and learning at all, and whether any effect depends on who makes the decision.

The TT format received a lower overall evaluation than the TS and SS formats. If this difference reflects the format itself rather than content or order, one approach is to improve the design of the TT format. An orientation activity before viewing that explains the format and provides an observation frame---for example, ``what to take away from a discussion between an expert and a facilitator''---may reduce the unfamiliarity of the format. Because the TT format may have left comparatively little room for learners' own reflection, a post-viewing reflection activity may also be useful, particularly for learners high in the RCE factor. Whether such measures improve the evaluation of the TT format remains to be tested.

In the free-response comments, some learners noted that the speakers' roles were unclear. Cognitive Load Theory distinguishes load arising from the complexity of the content itself (intrinsic load) from load imposed by the way information is presented (extraneous load)~\citep{sweller2019clt}. The load of inferring speaker roles arises from how the dialogue is presented, and may thus have functioned as extraneous load that did not directly contribute to content comprehension. When the roles or relationship between speakers are not self-evident from the dialogue itself, briefly disclosing speaker attributes before viewing (e.g., ``you are about to hear a conversation between two students'' or ``you are about to hear a discussion between an expert and a facilitator'') may help reduce this load. Making the speakers' roles explicit in advance may allow learners to focus their attention on understanding the dialogue content.

\subsection{Limitations}\label{sec:limitations}

This study has three limitations.

First, the outcome measures were brief, and learning outcomes were assessed only by self-report. Motivation was measured with an abbreviated scale consisting of one item per ARCS component, learning outcomes with two items, and overall evaluation with a single item. An abbreviated, one-item-per-component ARCS scale has also been used in prior work such as Watanabe and Kogo~\citeyearpar{watanabe2017jmooc}, but it is further simplified than the Reduced Instructional Materials Motivation Survey of Loorbach et al.~\citeyearpar{loorbach2015rimms}, which uses three items per component (12 items total). In addition, the two learning-outcome items---``I understood the main points of the lesson well'' and ``I think I could explain the main content of the lesson to a friend''---were self-reports, and no objective test, delayed post-test, or retention/transfer task was included. Self-report measures are susceptible to the influence of motivation, and learning outcomes may be overestimated for learners or in formats for which motivation is higher. Future research should measure each outcome with more items and examine the validity and reliability of these measures, and should combine objective tests with retention and transfer measures to examine whether the findings on learning outcomes extend to actual learning.

Second, this study was conducted in a single session (one day) with first-year students at a single public high school, which limits the generalizability and temporal scope of the findings. Replication across different school types, grade levels, subjects, and cultural contexts is needed, as are investigations of long-term learning effects (retention, transfer) and of changes with repeated viewing. Moreover, most participants had already experienced one dialogue-format TTS lesson approximately four to six months earlier in the study by Kumoi et al.~\citeyearpar{kumoi2026dialoguetts}. This experience may have reduced the novelty of the format or led to habituation to dialogue-format TTS, which may have affected the evaluation levels observed in this study. Because the main findings of this study concern the interaction with learner characteristics rather than absolute differences in evaluation across formats, and because this prior experience was shared by most participants, its impact on the interaction estimates may be limited. However, if its influence differed with learner characteristics---for example, with the CE factor, which reflects a tendency to seek challenging, new experiences---it may also have affected the interaction estimates.

Third, the three dialogue formats in this study each corresponded to different lesson content (Parts 1--3), and because all classes received the lessons simultaneously, counterbalancing the presentation order was not feasible; all participants viewed the lessons in the fixed order TS $\rightarrow$ SS $\rightarrow$ TT. Dialogue format, content, and presentation order therefore corresponded one-to-one, and their effects cannot be separated in this design. This applies not only to differences between formats but also to the interactions: the differences across formats in how learner characteristics were associated with the outcomes may reflect differences in content or order rather than in dialogue format. Future research should introduce an experimental design that orthogonalizes content and format---for example, by producing the same content in multiple formats---and should implement counterbalancing where the administration setting allows.

\section{Conclusions}

This study examined, within the ATI framework, the interaction between learner characteristics and dialogue format in LLM- and TTS-generated dialogue-based lessons. Using a repeated-measures design with 222 first-year high school students and linear mixed-effects models, we tested whether the associations of experiential learning style (CE and RCE factors) and critical thinking disposition with the three outcome measures (motivation, learning outcomes, and overall evaluation) differed across three dialogue formats (TS, SS, and TT).

For motivation, significant format $\times$ CE and format $\times$ RCE interactions were found: the association of the CE factor with motivation differed between the TS and TT formats, and learners higher in the RCE factor reported higher motivation in the TS and SS formats, with the slope of the RCE factor significantly larger in the TS format than in the TT format. This suggests, from an ATI perspective, that the relationship between learner characteristics and motivation can differ depending on the dialogue format. The format $\times$ critical thinking disposition interaction was not significant, but there was a positive main effect of critical thinking disposition, with higher disposition associated with higher motivation; however, this main effect was significant only under the Benjamini--Hochberg criterion. For learning outcomes, no format $\times$ learner characteristic interaction was detected, but there was a positive main effect of critical thinking disposition, with higher disposition associated with higher learning outcomes, and this main effect was significant under both correction methods. For overall evaluation, likewise, no format $\times$ learner characteristic interaction was detected, but unlike for learning outcomes, none of the main effects of learner characteristics was significant; only the main effect of format was significant, with the TT format rated lower than both the TS and SS formats. In other words, interactions between learner characteristics and format were detected only for motivation.

These findings offer implications for the design and further study of LLM- and TTS-generated dialogue-based lessons, although they should be treated as hypotheses because dialogue format was confounded with lesson content and presentation order in this study. First, the associations of the CE and RCE factors with motivation differed between the TS and TT formats, suggesting that dialogue format may be worth treating as a design variable together with learner characteristics. The present data, however, do not show whether adapting the format to learner characteristics improves motivation or learning. Future studies could test this by comparing a fixed-format condition with conditions in which the system assigns a format based on a brief pre-assessment of learner characteristics, the system recommends a format and learners make the final decision, or learners choose a format themselves. Second, the TT format received a lower overall evaluation than the TS and SS formats. If this difference reflects the format itself rather than content or order, an orientation that explains the format before viewing may reduce its unfamiliarity, and a post-viewing reflection activity may also be useful. Whether such measures improve the evaluation of the TT format also remains to be tested.

Beyond testing these implications, future research should address the limitations detailed in Section~\ref{sec:limitations}, in particular by producing the same content in multiple formats and counterbalancing the presentation order, so that the effects of dialogue format can be separated from those of content and order. LLM- and TTS-generated dialogue-based lessons remain a developing area, and the preliminary findings reported here add to the evidence on whether and how dialogue format should be taken into account when personalizing such lessons.

\authorcontributions{Conceptualization, F.W. and G.K.; methodology, F.W.; software, G.K.; formal analysis, F.W.; investigation, G.K.; data curation, F.W.; writing---original draft preparation, F.W.; writing---review and editing, F.W., G.K., T.S., T.I., Y.K., M.K. and S.H.; visualization, F.W.; supervision, G.K.; project administration, F.W.; funding acquisition, G.K. All authors have read and agreed to the published version of the manuscript.}

\funding{This work was supported by the JSPS Program for Forming Japan's Peak Research Universities (J-PEAKS), Grant Number JPJS00420240017.}

\institutionalreview{Ethical review and approval were waived for this study because it falls outside the scope of the ``Ethical Guidelines for Medical and Biological Research Involving Human Subjects'' (Notification No. 1 of 2021, Ministry of Education, Culture, Sports, Science and Technology; Ministry of Health, Labour and Welfare; and Ministry of Economy, Trade and Industry, Japan).}

\informedconsent{As described in the Materials and Methods section, the guardians had given comprehensive written consent to the school's provision of students' information to third-party organizations approved by the school principal, and the participating students were informed of the survey conditions at the beginning of the questionnaire; participation was voluntary, and only students who agreed to the stated conditions responded.}

\dataavailability{The data that support the findings of this study are not shared due to privacy and ethical considerations.}

\acknowledgments{The authors thank the high school students and teachers who participated in this study, as well as the individuals who provided valuable advice. This study used generative AI in two ways: to generate the lessons examined in this study (Claude 3.5 Sonnet, Anthropic, for the dialogue scripts; the Gemini TTS API, Google, for speech synthesis), as described in the Lessons section, and to assist in coding the free-response comments (Claude Sonnet 4.6, Anthropic), as described in the Materials and Methods section. During the preparation of this manuscript, the authors used Claude Sonnet 5 (Anthropic) for the initial translation of the manuscript from Japanese into English and Claude Opus 5.5 (Anthropic) to revise the English text. The authors have reviewed and edited the output and take full responsibility for the content of this publication.}

\conflictsofinterest{The authors declare no conflicts of interest.}

\reftitle{References}

\bibliography{references}

\end{document}